\def   \ni {\noindent}

\def   \ssk {\vskip  5truept}

\def   \bsk {\vskip 15truept}
 
\def   \newpage {\vfill\eject}
\def   \newline {\hfil\break}

\documentstyle[epsfig]{article}
\begin{document}

\hsize 5truein
\vsize 8truein
\font\abstract=cmr8
\font\keywords=cmr8
\font\caption=cmr8
\font\references=cmr8
\font\text=cmr10
\font\affiliation=cmssi10
\font\author=cmss10
\font\mc=cmss8
\font\title=cmssbx10 scaled\magstep2
\font\alcit=cmti7 scaled\magstephalf
\font\alcin=cmr6 
\font\ita=cmti8
\font\mma=cmr8
\def\ref{\par\noindent\hangindent 15pt}
\null


\title{\ni 
BEPPOSAX OBSERVATIONS OF THE GALACTIC SOURCE 
GS~1826-238 IN A HARD X-RAY HIGH STATE
}                                               

\bsk \bsk
\author{\ni S.~Del Sordo$^{1}$, F.~Frontera$^{2}$, E.~Pian$^{2}$,
S.~Piraino$^{1}$, T.~Oosterbroek$^{3}$, B.~A.~Harmon$^{4}$,
E.~Palazzi$^{2}$, M. Tavani$^{5}$, S.~N.~Zhang$^{4}$, and A. 
Parmar$^{3}$}

\bsk
\affiliation{$^{1}$ IFCAI-CNR, Via Ugo La Malfa 153, I-98146 Palermo, Italy}

\affiliation{$^{2}$ ITESRE-CNR, Via Gobetti 101, I-40129 Bologna, Italy}

\affiliation{$^{3}$ ESTEC, Astrophysics Division, Keplerlaan 1, 2200 AG
Noordwijk, The Netherlands}

\affiliation{$^{4}$ NASA Marshall Space Flight Center, Huntsville, AL 35812, USA}

\affiliation{$^{5}$ IFCTR, CNR, Via Bassini 15, I-20133 Milano, Italy}
                                                
\bsk
\baselineskip = 12pt

\abstract{ABSTRACT \ni

The BeppoSAX Narrow Field Instruments observed the galactic source GS~1826-238
in October 1997, following a hard X-ray burst with a peak flux of about 100
mCrab detected by BATSE.  Two short X-ray bursts ($\sim$150 seconds) were
detected up to 60 keV, with larger amplitude and duration at lower energies (up
to a factor 20 times the persistent emission). This confirms the proposed
identification of the source as a weakly magnetized neutron star in a LMXRB
system. For both persistent and burst states, the spectrum in the 0.4-100 keV
energy range is well fitted by an absorbed black body, plus a flat power-law
($\Gamma \sim 1.7$) with an exponential cutoff at 50 keV.

}
\bsk
\baselineskip = 12pt

\bsk
\baselineskip = 12pt


\text{\ni 1. INTRODUCTION
\ssk
\ni     

GS~1826-238 was serendipitously discovered by the {\it Ginga} LAC in September
1988 [1] with a flux of 26 mCrab and a hard power-law spectrum (photon index
$\Gamma = 1.7$), and optically identified with a $V \simeq 19.3$ star [2].
Observations both a month before and after the discovery by the {\it Ginga} ASM,
by TTM in 1989 and by ROSAT in 1990 and 1992 found comparable flux levels [2,3]. 
In 1994 the source was detected at a 7$\sigma$ level with OSSE above 50 keV [4]
with a steep power-law spectrum ($\Gamma = 3.1$).  

Due to its flickering flux variability and hard X-ray spectrum, reminiscent
of the behavior of Cyg X-1, the source was tentatively classified as a black
hole candidate.  Three X-ray bursts detected on 31 March 1997 with the WFC
onboard BeppoSAX [5], and two optical bursts [6] suggest instead that the
system contains a lowly magnetized neutron star.  We
focus here on the
spectral shape of the source during intense, hard X-ray states and on the
spectral evolution exhibited in rapid X-ray bursts.

\begin{figure}
\vspace{-2.5cm}
\centerline{\psfig{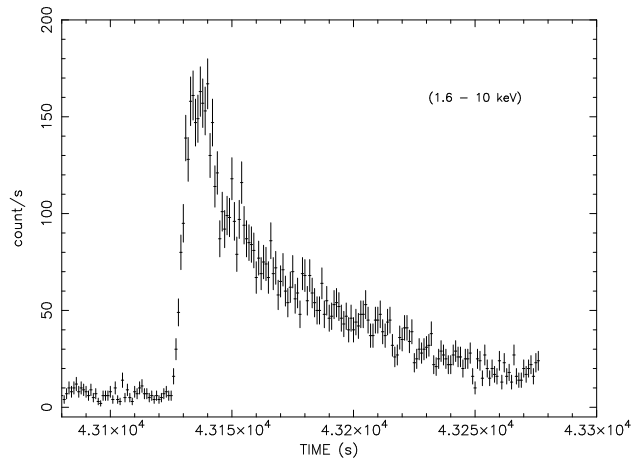}}
\vspace{0.3cm}
\caption{FIGURE 1. Portion of the MECS Light curve during the second
pointing binned with a 
1 second resolution. The start time of the burst is 1997 October 18.12 UT.
}
\end{figure}


\newpage
\bsk
\ni 2. OBSERVATIONS, DATA ANALYSIS AND RESULTS
\ssk
\ni 

Observations with the BeppoSAX Narrow Field Instruments took place on 6-7,
17-18, and 25-26 October 1997 within a Target of Opportunity program for the
monitoring of X-ray transients in active state in hard X-rays.  The first
pointing was triggered by the BATSE detection of a hard X-ray signal with
a peak
flux of about 100 mCrab. On source MECS exposures lasted typically for 25-30
ks, while LECS and PDS exposures were typically 30\% and 50\% shorter,
respectively.  The target was significantly detected with the PDS up to 100
keV. Data reduction was performed following standard methods (see,
e.g., [7]).
The MECS light curves exhibit two pronounced X-ray bursts of a factor 20
amplitude in the range 1.6-10 keV and $\sim$150 seconds duration on October
18.12 (Fig. 1) and 26.19 UT. (Due to instrumental visibility constraints,
the LECS only partially detected the first burst.) Both bursts were also
detected, up to 60 keV, by the PDS. The flux exhibits a linear rise, which
appears faster at higher energies, followed by an exponential decay.  Burst
amplitudes and durations decrease with increasing energy.  Excluding the
bursts, no significant variability is seen at any frequency within a same
pointing, nor from epoch to epoch.  
For the spectral analysis, the LECS, MECS, HPGSPC and PDS data have been
considered in the ranges 0.4-6 keV, 1.6-10 keV, 6-30 keV, and 15-100 keV
(15-50 keV for the bursts), respectively.  Both persistent and burst
emission spectra cannot be fitted by an absorbed black body model only, but
they exhibit a high energy tail which is accounted for by a power-law model
plus a high energy cutoff (Fig.  2).  For the October 17-18 observation, we
obtain fitted temperatures of $T_{BB} = 0.94 \pm 0.05$ keV and $T_{BB} = 1.9
\pm 0.1$ keV for persistent and burst emission, respectively, power-law
photon indices $\Gamma = 1.34 \pm 0.04$ and $\Gamma = 1.1 \pm 0.4$, and
exponential cutoffs $E = 49 \pm 3$ keV and $E = 12 \pm 4$ keV.  The last
value is very uncertain, due to the more limited energy range used for the
fit.  A fit to time-resolved spectra in the 2-10 keV range 
along the burst light curve
indicates
that the temperature decreases during the burst decay.  The burst emission
spectrum has been fitted without subtracting the underlying persistent
signal, to avoid neglecting the possible influence of the burst on the
accretion flow (see [8], and references therein).

\begin{figure}
\vspace{-3cm}
\centerline{\psfig{file=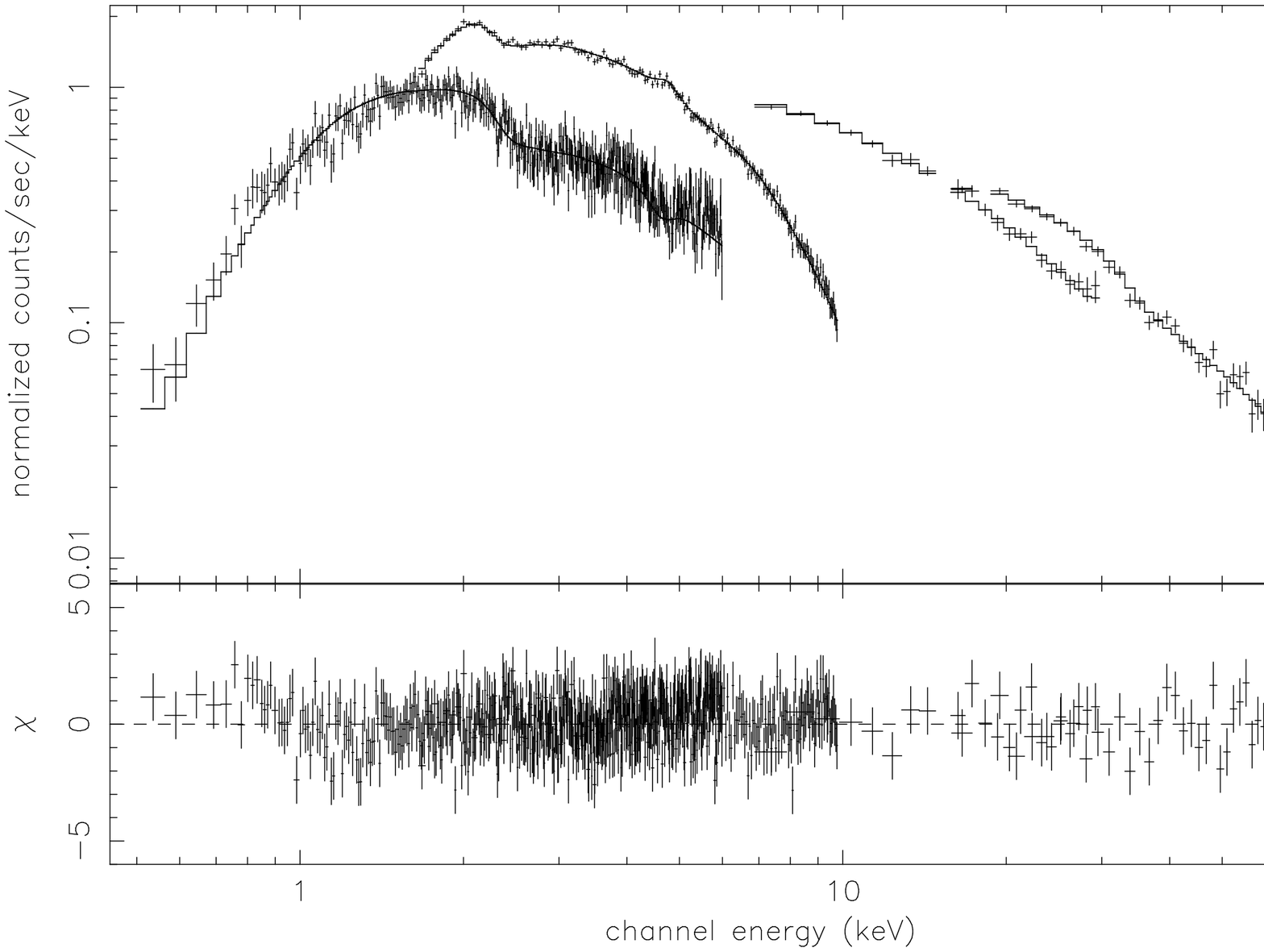, width=8cm}}
\vspace{-3cm}
\centerline{\psfig{file=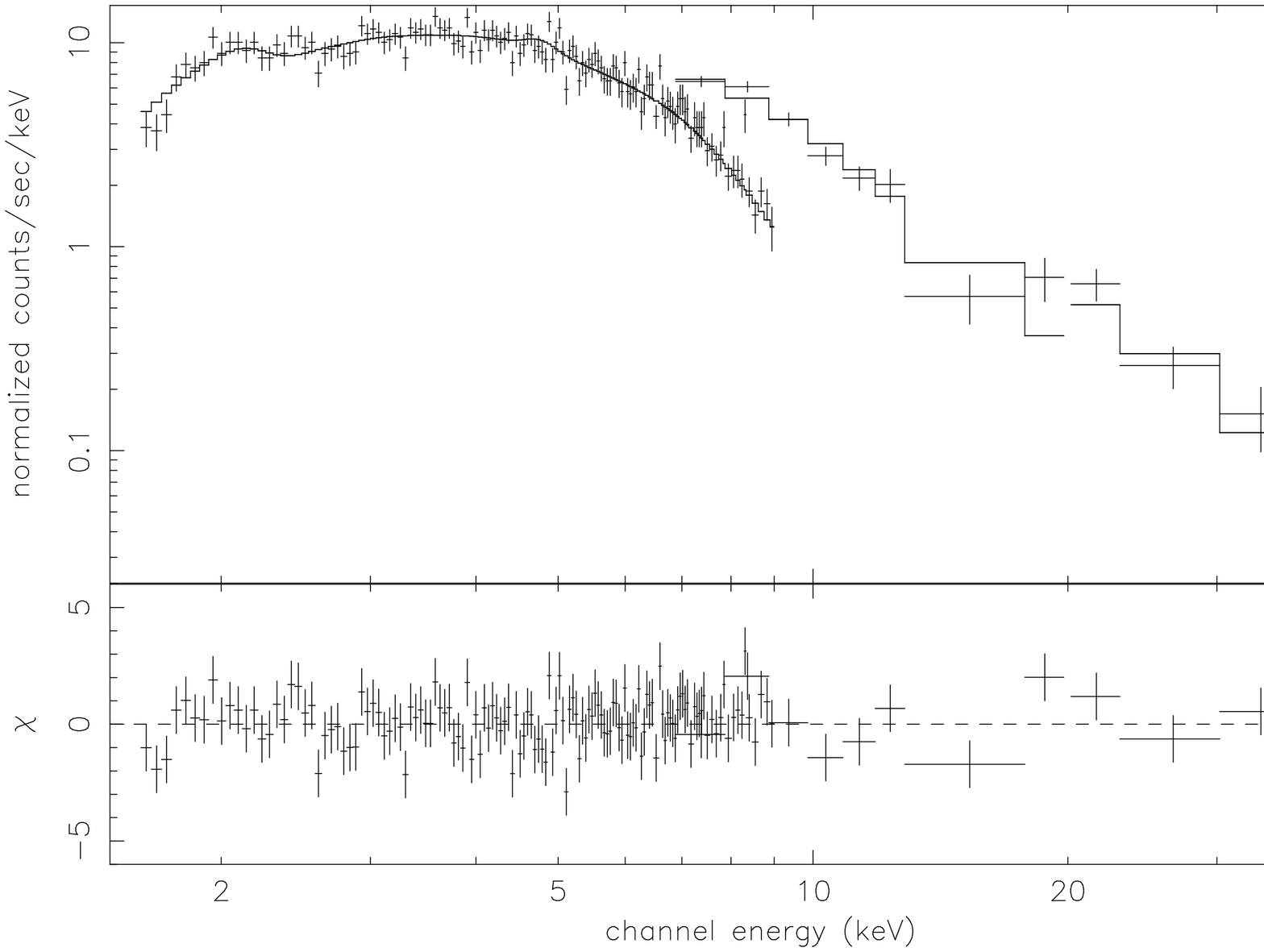, width=8cm}}
\vspace{0.3cm}

\caption{FIGURE 2. Top panel: spectrum of the persistent emission measured on
October 17 by the four Narrow Field Instruments along with the model which best
fits the overall spectrum.  The average flux level in the 2-10 keV band is $5.6
\times 10^{-10}$ erg s$^{-1}$ cm$^{-2}$. The best fit value for the neutral
hydrogen column density, $N_{\rm H} \simeq 4.7 \times 10^{21}$ cm$^{-2}$, is
consistent with the Galactic one. The residuals with respect to $\chi^2 = 1$
are
also shown.  Bottom panel: spectrum of the burst+persistent emission measured by
the MECS, HPGSPC and PDS during the October 18 event.  Best fit model curve and
residuals are also shown.

}
\end{figure}


\bsk
\ni
3. DISCUSSION
\ssk
\ni

The energy-dependent amplitude of the bursts detected by BeppoSAX and their
shorter duration at higher energies strongly suggest a  thermonuclear
origin.  This is confirmed by the good fit obtained with a thermal model up
to 10 keV and by the decreasing black body temperature during the burst
decay. The present and previous observations of burst activity in
GS~1826-238 rule out the black hole nature of the central accretor and
suggest the presence of a weakly magnetized neutron star in a LMXRB system. 
While the temperature of the black body doubles during the burst with
respect to the persistent emission, the slope of the high energy power-law
does not significantly vary, although its normalization is a factor of two 
larger in burst than in persistent state, indicating that the hard tail is
intrinsically significant in the burst state too.  A similar finding was
previously reported for X1608--52 by Nakamura et al. [9], although the
explored spectral range was more limited.  Our fitted temperature of $\sim$2
keV during the burst is consistent with the inverse correlation Nakamura
et al.  find between temperature and hard tail intensity (see their Table
2).  This high energy component is tentatively interpreted in terms of
inverse Compton scattering of soft black body photons off high energy
electrons in a hot region surrounding the neutron star (see however Day \&
Done [10]).  
The break at 50
keV, reminiscent of that observed by BATSE in X1608--52 [11], indicates
that, unlike in black hole candidates, the neutron star surface emission at
soft energies tends to cool the hot Comptonization region.  The
temporal spacing of the two bursts detected by the present BeppoSAX
observations is consistent, within the 40 minutes uncertainty, with the 5.67
hours period found by Ubertini et al.  [12] during 2.5 years of WFC
observations.

}

%

\bsk
\baselineskip = 12pt


{\references \ni REFERENCES
\ssk

[1] Makino, F. 1988, IAU Circ. No. 4653;
[2] Barret, D., Motch, C., \& Pietsch, W. 1995, A\&A, 303, 526;
[3] in't Zand, J. J. M. 1992, PhD Thesis, University of Utrecht;
[4] Strickman, M., et al. 1996, A\&AS, 120, 217;
[5] Ubertini, P., et al. 1997, IAU Circ. No. 6611;
[6] Homer, L., Charles, P. A., \& O'Donoghue, D. 1998, MNRAS, 298, 497;
[7] Chiappetti, L., et al. 1998, Nuc.  Phys. B (Proc. Suppl.)
69/1-3, 340;
[8] Lewin, W. H. G., Van Paradijs, J., \& Taam, R. E. 1995, in X-ray
Binaries, eds. W. H. G. Lewin, J. Van Paradijs, and E. P. J. Van den Heuvel,
Cambridge Astrophysics Series;
[9] Nakamura, N., et al. 1989, PASJ, 41, 617;
[10] Day, C. S. R., \& Done, C. 1991, MNRAS, 253, 35P;
[11] Zhang, S.N., et al. 1996, A\&AS, 120C, 279;
[12] Ubertini, P., et al. 1998, this Conference.

}                      

\end{document}